\documentclass{sig-alternate}
\usepackage{microtype}
\usepackage[utf8]{inputenc}
\usepackage{graphicx}
\usepackage{subcaption}
\usepackage{tabulary}
\usepackage{booktabs}
\usepackage{url}


\usepackage[unicode=true]{hyperref}
\hypersetup{breaklinks=true,
            bookmarks=true,
            pdfauthor={Anonymous},
            pdftitle={Virtual Portraits},
            colorlinks=false,
            urlcolor=blue,
            linkcolor=magenta,
            pdfborder={0 0 0}}
            
\newcommand{\spara}[1]{\smallskip\noindent{\bf #1}}

\numberofauthors{3}

\author{
\alignauthor
Eduardo Graells-Garrido \\
Universitat Pompeu Fabra \\
Barcelona, Spain \\
\texttt{eduard.graells@upf.edu}
\alignauthor
Mounia Lalmas \\
Yahoo Labs \\
London, UK \\
\texttt{mounia@acm.org}
\alignauthor
Daniele Quercia \\
Yahoo Labs \\
Barcelona, Spain \\
\texttt{dquercia@yahoo-inc.com}
}

\global\boilerplate={Unpublished manuscript. Ask authors before citing/distributing.}
\global\copyrightetc{November 2013}

\title{Data Portraits: Connecting People of Opposing Views}
\begin{document}

\maketitle

\begin{abstract}
Social networks allow people to connect with each other and have conversations on a wide variety of topics. 
However, users tend to connect with like-minded people and read agreeable information, a behavior that leads to group polarization.
Motivated by this scenario, we study how to take advantage of partial homophily to suggest agreeable content to users authored by people with opposite views on sensitive issues. 
We introduce a paradigm to present a data portrait of users, in which their characterizing topics are visualized and their corresponding tweets are displayed using an organic design. 
Among their tweets we inject recommended tweets from other people considering their views on sensitive issues in addition to topical relevance, indirectly motivating connections between dissimilar people. 
To evaluate our approach, we present a case study on Twitter about a sensitive topic in Chile, where we estimate user stances for regular people and find intermediary topics.
We then evaluated our design in a user study. We found that recommending topically relevant content from authors with opposite views in a baseline interface had a negative emotional effect. We saw that our organic visualization design reverts that effect. 
We also observed significant individual differences linked to evaluation of recommendations. 
Our results suggest that organic visualization may revert the negative effects of providing potentially sensitive content.
\end{abstract}




\section{Introduction}

Today, with social networks, users have less barriers to communicate. Geographical distance is no longer a limitation to interact with others or to know what is happening anywhere in the world. In particular, real-time streams keep people aware of what is happening simply by reading content posted from other accounts. 
Discussions for instance about political events are driven by the usage of hashtags and anyone can contribute and participate without being connected to other participants in the discussions. 
However, is it really that good? 

Social research has shown that people tend to connect and interact mostly with people with similar beliefs, a phenomena known as \emph{homophily}. In addition, users prefer to access agreeable information instead of information that challenges their beliefs. These problems are augmented when social networks recommend content based on what users already know, on what are their current connections, and what people that are similar to them have done and liked before. This phenomenon of being surrounded only by similar people and having access to agreeable or likeable information is known as the \emph{filter bubble} \cite{pariser2011filter}. 

As a result, groups of users that share different points of view tend to polarize and disconnect from other groups. 
Previous work has focused on how to motivate users to read challenging information or how to motivate a change in behavior. This ``direct'' approach has not been effective as users do not seem to value diversity or do not feel satisfied with it, a result explained by \emph{cognitive dissonance} \cite{festinger1962theory}, a state of discomfort that affects persons confronted with conflicting ideas, beliefs, values or emotional reactions. 
Because social networks make heavy use of recommendations, both curated by other peers and machine generated, the motivation behind our research is: \emph{can we take advantage of user engagement with recommendations to \emph{indirectly} promote connection to people of opposing views?}

Our methodology recommends content to users through a \emph{data portrait} \cite{donath2010data} of themselves, making them aware of their characterizing topics, concepts and relations in a social network. We call this characterization \emph{user preferences}. User preferences are displayed following an organic design, based on familiar elements (wordclouds) and new stimuli (circles in organic layouts). 
Then, we take advantage of partial homophily as we inject recommended content in the context of the preferences of the portrayed user.
Besides topical relevance, the recommended content considers an additional factor: the difference in stances on sensitive issues for both \emph{recommender} and \emph{recommendee}. We call this difference \emph{view gap}. Hence, we nudge users to read content from people who may have opposite views, or high view gaps, in those issues, while still being relevant according to their preferences.
The thinking behind this ``indirect'' recommendation is that if a connection is stablished, it could be possible that the portrayed user will receive potentially challenging content in the future. This content could be better tolerated because of the \emph{primacy effect} in impression formation \cite{asch1946forming}, or, as the popular interpretation says: \emph{first impressions matter}. 
This paper is a first step in this direction by presenting an organic approach to the visual design of data portraits.

To evaluate our methodology, we performed a case study on the social network Twitter\footnote{\url{https://twitter.com}, visited 02-October-2013.}  where the sensitive issue is abortion and the user population is from Chile. 
Twitter is a social network that contains a high percentage of public discussions on what is currently happening. 
Chile has one of the strictest abortion laws in the world \cite{abortionChile}, and discussion on social networks is polarized in particular now ahead of upcoming presidential elections (to be held in November 2013). 
Twitter also provides two types of recommendations: the \emph{who to follow?} widget \cite{gupta2013wtf}, 
and \emph{\#Discover}, a sub-page that contains recent tweets published, retweeted or favorited by others. All of these recommendations are driven by user connectivity and user activity. 
Making recommendations on Twitter is not new, and we exploit this artefact to make indirect recommendations using our data portrait paradigm.

Using tweets crawled during July and August 2013, we use our methodology to establish an abortion stance for any Twitter account from Chile as a linear combination of the two opposite stances: \emph{pro-choice} and \emph{pro-life}. From all accounts that have \emph{tweeted} about abortion, we select a sample of \emph{regular people}. We find that indeed there is topical diversity; people of opposite views discuss many other topics, and some common to both stances. 
Then, we performed a user study where participants were divided into three groups: a baseline, where the data portrait consisted of an interactive wordcloud and two standard list of tweets: one from the portrayed users, and other with recommendations considering user preferences only; a treatment I, where the recommendations considered user preferences and view gaps; and a treatment II, where the data portrait is based on our organic design and the recommendations considered user preferences and view gaps.
Our results indicate that the user experience with our novel data portrait design is comparable to a familiar baseline. The injection of content from people with opposite views had a negative effect on the emotional reaction of users when comparing the baseline and the first treatment. However, our organic design helped to recover the emotional reaction to its previous levels. We also found how individual differences are related to recommendation attributes: openness is linked to perceived interestingness, and engagement with our design is related to perceived serendipity.

This paper makes the following contributions:
(1) An organic visualization design that depicts a virtual portrait of a user based on its characterizing content; 
(2) A methodology to characterize users from social networks into a set of topics and stance on sensitive issues, and recommend tweets based on both;
(3) A case study of a sensitive issue and a corresponding user study, demonstrating the potential of our chosen methodologies and visualization design.

\section{Background}

\spara{Visualization and Data Portraits}
Visualizations of social network data cover a wide range of applications, such as event monitoring \cite{dork2010visual,marcus2011twitinfo}, visual analysis \cite{diakopoulos2010diamonds}, group content analysis \cite{archambault2011themecrowds}, information diffusion \cite{viegas2013google+} and ego-networks \cite{2005-vizster}. Our work is related to the field known as \emph{Casual Information Visualization}, defined as \emph{``the use of computer mediated tools to depict personally meaningful information in visual ways that support everyday users in both everyday work and non-work situations''} \cite{pousman2007casual}. The focus on everyday situations imply that there does not need to be a concrete task to be completed.
To visualize user profiles we build \emph{data portraits} \cite{donath2010data}, which are ``abstract representations of users' interaction history'' \cite{xiong1999peoplegarden}. These portraits have been built using content from e-mail \cite{viegas2006visualizing}, personal informatics systems \cite{assogba2009mycrocosm}, Twitter profiles \cite{lexigraphs} and internet forums \cite{xiong1999peoplegarden}. 
Since we focus on the topical aspect of profiles, we make use visual depictions of collections of words known as \emph{wordclouds} \cite{viegas2008timelines}.

\spara{Selective Exposure.}
Selective exposure refers to the tendency of people to seek for information that reinforces their current beliefs and ideas. Exposure to challenging information causes \emph{cognitive dissonance} \cite{festinger1962theory}, an uncomfortable state of mind that individuals try to alleviate by discarding the information or avoiding it. Previous works have focused on how to minimize or avoid this dissonance to improve exposure to challenging information.
In \emph{NewsCube} \cite{park2009newscube} several automatically determined aspects of news stories are presented to mitigate the problem of media bias and allow users to access diverse points of views.
A study on sorting and highlighting of political opinions identified two groups of users: \emph{diversity-seeking} and \emph{challenge-averse} \cite{munson2010presenting}, where the latter was shown to be equally satisfied with a list of mostly non highlighter agreeable items and a couple of highlighted challenging items, and a list of only agreeable items.
In \emph{OpinionSpace} \cite{faridani2010opinion}, users were presented with different visualizations of opinions.
The usage of a visual approach did not reduce selective exposure, although it generated more engagement than a baseline and its users were more respectful with those having opposite opinions.
In \cite{munson2012} users were presented with a cartoon representation of the equilibrium of their reading behavior in terms of political diversity: reading only articles from one political side would make the figure in the cartoon almost falls, whereas reading a fair proportion of political articles from both sides would keep the figure balanced.
Finally, \cite{liao2013beyond} reported that information seekers preferred agreeable information even when having to choose between agreeable and challenging items presented side by side, and high topic involvement and perceived threat were found to be influencing factors in selective exposure.

While previous works refer to direct presentation and access to diverse content, our work takes a different, indirect approach. We do not explicitly display challenging content for two reasons: \emph{1)} users often do not value diversity \cite{munson2010presenting}, and \emph{2)} we take advantage of the \emph{primacy effect} on impression formation \cite{asch1946forming}. 
Hence, we do not assess selective exposure directly, as our goal is to connect people of opposing views on a sensitive topic, but having non challenging initial interactions.

\spara{User Classification and Characterization.}
We consider two levels at which users can be modeled.
At the high level,  self-reported location, name and biography can be used to detect user demographic features \cite{mislove2011understanding}, which, in addition to linguistic features of user generated content, user behavior and user connectivity, can be used to classify users into political parties (and other categorizations) using machine learning algorithms \cite{pennacchiotti2011democrats}. 
However, because regular people may not be as vocal as politically involved people, the accuracy of the prediction algorithms is often over-estimated \cite{ICWSM136128}. 
At the low level, using a knowledge base allows the detection of entities mentioned in tweets which, in turn, allows to assign characterizing categories to a Twitter profile \cite{michelson2010discovering}. Topic modelling \cite{ramage2010characterizing} is used to calculate latent topics from a collection of microblogs that then characterizes the collection. Then, user profiles can be characterized by the probability distributions of how those topics contribute to their content. A dimensionality reduction approach is proposed by OpinionSpace \cite{faridani2010opinion}, where users are asked for their opinion of several topics to build an \emph{``opinion profile''}.

In our work, user views in sensitive issues are estimated using a mixture of a linguistic approach and opinion profiles \cite{faridani2010opinion}. Since our work is focused on visualization, this is sufficient to obtain a good enough stance determination, as a linguistic approach has been shown to be reliable \cite{pennacchiotti2011democrats}. For user preferences or topical characterization we work with \emph{n-grams} to detect characteristic word sequences such as place names, artists, movies, and other lexical features. 
We recommend tweets based on topical relevance and the differences in user views. Although state-of-the-art algorithms also consider social relations and different measures of tweet quality \cite{chen2012collaborative} these features are expensive to estimate, which is outside the scope of this paper.

\spara{Sensitive Issues and Politics in Social Media.}
Opposite communities in social media have been studied. 
For instance,  \cite{adamic2005political} found that liberal and conservative blogs in the US linked mostly to blogs of the same political party. %
\cite{yom2012pro} studied the interaction between two opposite groups of users on Flickr (a social network about photography)\footnote{\url{http://flickr.com}, accessed on 08-October-2013.}, one related to \emph{pro-anorexia} and the other to \emph{pro-recovery}. Pro-recovery users injected content into pro-anorexia groups but the intervention was counterproductive.
The injection of political content on Twitter from opposite parties has been studied in \cite{ICWSM112847}, and was shown to motivate interactions between politically vocal users. However, although the \emph{mention network} was found to become less polarized, the \emph{retweet network} was not. Debates on abortion between \emph{pro-life} and \emph{pro-choice} advocates have been studied in \cite{yardi2010dynamic}, who showed that the interaction between persons having the same stance reinforced group identity, and discussions with members of the opposite group were found to be not meaningful, partly because the interface did not help in that aspect.
As noted in \cite{yardi2010dynamic}, people hardly changed position on abortion based on discussions on Twitter. However, being connected to a more diverse group of people may help create a more meaningful discussion and affect people points of view instead of merely reinforcing them. This is what we strive for in our work.

\begin{figure*}[thb]
\centering
\includegraphics[width=0.85\textwidth]{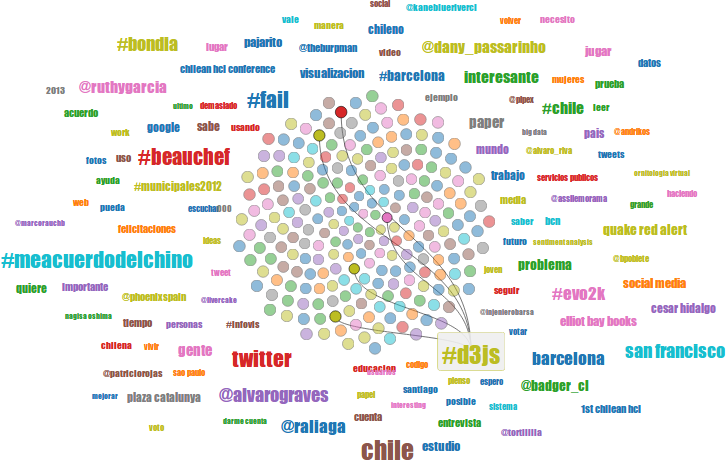}
\caption{Our data portrait design, based on a wordcloud and an organic layout of circles. The wordcloud contains characterizing topics and each circle is a tweet about one or more of those topics.  Here, the user has clicked on her or his characterizing topic \texttt{\#d3js} and links to corresponding tweets have been drawn.}
\label{fig:application}
\end{figure*}

\section{Data Portrait}

%

Our data portrait paradigm is aimed at creating a ``global image'' of a Twitter account, an image that depends on what users want their account to project to others. Existing work on data portraits have focused either on disruptive and artistic designs, or solely analytical ones. We acknowledge that regular users might exhibit resistance when faced to disruptive changes in the interfaces they are used to. Therefore, as shown in Figure \ref{fig:application}, we propose to add  a new stimuli to a familiar element, \emph{wordclouds}, which changes in a substantial way the interaction with it, while providing a friendly and evocative appearance based on organic patterns.

\spara{Depicting User Topics.}
In data portraits, text is important as \emph{``it provides immediate context and detail''} \cite{donath2010data}. We make use of the topics that characterize a Twitter account as input for a wordcloud in each portrait. Wordclouds exist since 1997 \cite{viegas2008timelines} so we expect users to be familiar with them. The font-size of each topic varies according to the score given to the topic: a higher score implies a higher font-size. Word positioning and color are random, as in popular wordclouds \cite{viegas2009participatory}. 
To make each topic clickable like a button, we position topics in the canvas using bounding box collision detection to minimize overlap. We extend each bounding box along its axes to have a larger clickable area.

\spara{Depicting Tweets.}
Tweets are first depicted as circles in the middle of the wordcloud, as we want to insinuate that tweets are originating those topics. The layout organizing the circles is based on an organic model of pattern of florets from Vogel \cite{vogel1979better}, defined in polar coordinates as:
$$
r = c \sqrt{n}  \text{ and }
\theta = n \times 137.508
$$
where $c$ is a constant that defines how separated circles are, $137.508$ is the \emph{golden ratio} and $n$ is the reverse chronological ordered index of each tweet (the oldest tweet will have index $1$). The color of each circle is based on the color assigned to the primary topic of the corresponding tweet. By using collision detection, wordcloud elements do not overlap with the circles while positioning the topics in the canvas.

\begin{figure}[t]
\centering
\includegraphics[width=0.33\textwidth]{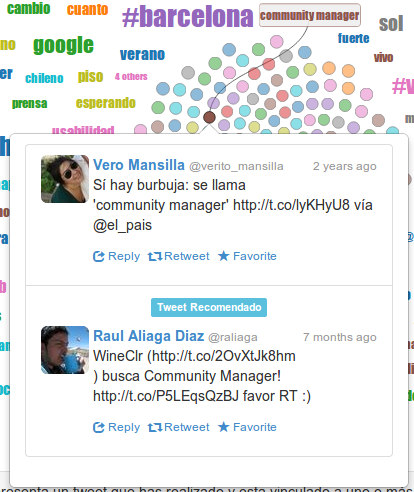}
\caption{Display of tweets inside a pop-up \emph{speech balloon}.}
\label{fig:vis_recommendation}
\end{figure}

\spara{Context, Nudging and Interaction.}
When  a user selects a tweet, its content is presented inside a \emph{speech balloon} designed with a format that resembles the native format in Twitter. The tweet includes native options such as \emph{retweet}, \emph{reply} and \emph{mark as favorite}. 
To nudge interaction with others, in addition to the portrayed individual tweet, we display \emph{related tweets} made by unknown individuals, as shown in Figure \ref{fig:vis_recommendation}. 
To provide topical context, tweets and topics are connected on-demand with \emph{bezier curves}, as shown in Figure \ref{fig:application}. 
It is possible that one topic is linked to many tweets; and one tweet to many topics. Clicking on a topic highlights the topic, giving it a button-like appearance and displaying the connection curves to the corresponding tweets. Clicking on a circle displays the corresponding tweet and the connection curves to the corresponding topics. Closing the tweet does not discard the connection curves, hence, users can explore the topics based on how they are interconnected according to the tweets. The curves are discarded by selecting another topic or by clicking into empty space in the canvas.

\section{Methodology}

We consider the social media site of Twitter, where users publish micro-posts known as \emph{tweets}, each having a maximum length of 140 characters. Each user is able to \emph{follow} other users and their tweets. We consider tweets about sensitive issues, which can be defined as the topics for which the stances or opinions tend to divide people. For instance, abortion is a sensitive issue in many countries, whereas musical tastes is usually not (even if people have different tastes in music). Often users annotate their tweets with \emph{hashtags}, which are text identifiers that start with the character \texttt{\#}. For instance, \texttt{\#prochoice} and \texttt{\#prolife} are two hashtags related to two abortion stances. 

Our aim is to build a tool that recommends users tweets that they may like, and unknown to them, that come from people who hold opposite views in a particular sensitive issue (in our case abortion for users in Chile). 
We thus need to determine for each user, what is his or her view with respect to the sensitive issue under consideration, and what he and she is interested in general (sport, dining, film, etc.).

\spara{User Views}
For any sensitive issue under consideration, we collect relevant tweets. Relevance is determined based on keywords and hashtags contained in tweets. The collected tweets fall into one of the several issue stances. For each stance, we compute a \emph{stance vector} in which each dimension refers to the importance of a given word taken from the tweets containing that word. This importance is calculated using the TF-IDF weighting scheme used in the vector space model in information retrieval \cite[Chapter 3]{baeza2011modern}.
We also define a \emph{user vector}, in which each dimension refers to the importance of a given word taken from the user tweets. Finally, we define the \emph{user stance} as a vector where each dimension corresponds to the cosine similarity between the user vector and the several issue stance vectors. For a pair of users, we define their \emph{view gap} as the distance between their stances.

\spara{User Preferences} 
To obtain the \emph{top-k} topics that characterize a user (Twitter account), we compute the most common n-grams (with $n$ up to $3$) of the user tweets. Each set of n-grams is ranked separately, where the score given to each topic is a linear combination of the number of occurrences (frequency) and how long ago has the n-gram been written (time). The weights for frequency and time depend on the number of words in the n-gram, as unigrams are expected to be more frequent than bi- or tri-grams. We then combine the n-gram sets and select the top ranked ones. We do this for three types of content: \emph{mentions} (interactions), \emph{hashtags} (explicit topics) and \emph{words} (implicit topics).

\spara{Tweet Recommendations}
Having each user views and preferences, we are now able to suggest tweets to users. We do so by ranking tweets based on both:
\begin{itemize}\itemsep0pt
	\item \emph{Topical Relevance}: whether they match the user preferences. To estimate topical relevance we use the cosine similarity between the user preferences under consideration and a pool of candidate tweets.
	\item \emph{Opposite Views}: whether their authors have a considerable view gap with the user under consideration.  
\end{itemize}
In this way, suggested tweets are still relevant but come from a diverse pool of users, where diversity is with respect to user stances on a topic.

\section{Case Study: Abortion in Chile}
\label{section:case_study}

One of the most sensitive issues in Chile is abortion. Chile has one of the most severe abortion laws in the world; abortion is illegal and there is no exception. The history of abortion in Chile is long, being declared legal in 1931 and illegal again in 1989. Given that the presidential elections will be held soon and the occurrence of several protests around themes such as abortion, public education and gay marriage, social networks are being used to spread  ideas and generate debates. In this section we describe how we represent \emph{user views} on abortion for a set of ``regular'' users from Chile in Twitter. We then describe how we determine the topical diversity for those users, to showcase that we can find relevant content to recommend to them according to their \emph{user preferences}.

\spara{Dataset}
We crawled tweets for the period of July to August 2013 using the \emph{Twitter Streaming API}.\footnote{\url{https://dev.twitter.com/docs/streaming-api}, accessed on 3-October-2013. The data was crawled by the first author.} Table \ref{table:information_space} shows a summary of the crawled data.

\begin{table}[t]
\centering
{\scriptsize
\begin{tabular}{lc}
  \toprule
  Data & \#  \\
  \midrule
  Tweets & $3002299$\\
  Accounts & $798590$ \\
  Keywords & $1289901$ \\
  \midrule 
  Accounts with Abortion-related Tweets & $40201$ \\
  \bottomrule
\end{tabular}}
\caption{Data crawled from Twitter during July and August 2013.}
\label{table:information_space}
\end{table}

\subsection{User Views}
To represent user views, we look at the most discussed sensitive issues on Twitter. To find those issues, we crawled tweets in July using keywords for known issues and other keywords and hashtags for Chile (location names and presidential candidate names). From those tweets, we manually inspected the most used keywords to find the top-100 related to sensitive issues. Then, in August, we crawled tweets using those keywords, plus emergent keywords and hashtags from unexpected news and events related to these issues. 
Finally, from those tweets we selected the final top-100 keywords to define a corpus of documents built with the tweets containing them. 
We built two \emph{stance vectors}: \emph{pro-life} and \emph{pro-choice}. The keywords used to build both vectors are specified in Table \ref{table:abortion_documents}. The importance of each word in each stance vector is weighted according to its TF-IDF with respect to a previously defined corpus.

\begin{table*}[t]
\centering
\scriptsize
\begin{tabulary}{\textwidth}{L|L}
\toprule
Stance & Keywords  \\
\midrule
\emph{Pro-choice} & \texttt{\#abortolibre}, \texttt{\#yoabortoel25}, \texttt{\#abortolegal}, \texttt{\#yoaborto}, \texttt{\#abortoterapeutico}, \texttt{\#proaborto}, \texttt{\#abortolibresegurogratuito}, \texttt{\#despenalizaciondelaborto} \\
\midrule
\emph{Pro-life} & \texttt{\#provida}, \texttt{\#profamilia}, \texttt{\#abortoesviolencia}, \texttt{\#noalaborto}, \texttt{\#prolife}, \texttt{\#sialavida}, \texttt{\#dejalolatir}, \texttt{\#siempreporlavida}, \texttt{provida}, \texttt{\#nuncaaceptaremoselaborto} \\
\midrule
General & \texttt{\#marchaabortolegal}, \texttt{aborto}, \texttt{abortistas}, \texttt{abortista}, \texttt{abortos}, \texttt{abortados}, \texttt{abortivo}, \texttt{\#bonoaborto} \\
\bottomrule
\end{tabulary}
\caption{Keywords used to characterize the pro-choice and pro-life stances on abortion. General keywords plus stance keywords were used to find people who talked about abortion in Twitter.}
\label{table:abortion_documents}
\end{table*}

\begin{figure}[t]
\centering
\includegraphics[width=0.32\textwidth]{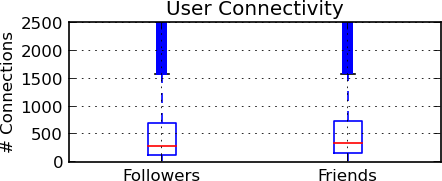}
\caption{Boxplot showing distribution of in-degree (\emph{followers}) and out-degree (\emph{friends}) of chilean users who have tweeted about abortion.}
\label{fig:case_study_boxplot}
\end{figure}

In our data, the number of users who published tweets using keywords related to abortion (see Table \ref{table:abortion_documents}) is $40,197$. Because we wanted to analyze \emph{regular users} (and not accounts that neither interact with others nor participate into discussions), we used a boxplot to identify outliers according to connectivity (see Figure \ref{fig:case_study_boxplot}). The median of \emph{followers} (in-degree) is $283$ and the maximum non-outlier value is $1,570$. The median of \emph{friends} (out-degree) is $332$ and the maximum non-outlier value is $1580$. We filtered users according to their in- and out-degrees. Then, we filtered users who had less than $1,000$ tweets. Finally, we selected users that reported their location as Chile in their profiles and that have a known gender. To geolocate users according to their self-reported location we used a methodology from \cite{graells2013santiago}, and to detect gender we used lists of names. After applying those filters we had a candidate pool of $3,350$ users. For these candidates we downloaded their latest $1,000$ tweets and estimated their \emph{user vectors}, by weighting the words used in their tweets according to our corpus of sensitive issue keywords.

\begin{figure}[t]
\centering
\includegraphics[width=0.32\textwidth]{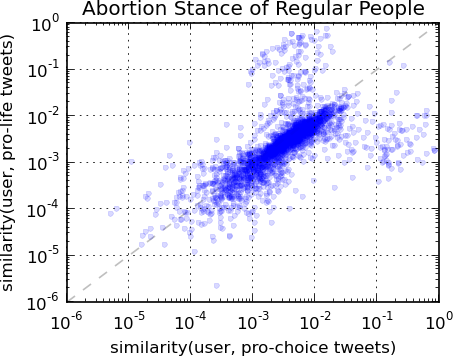}
\caption{Distribution of user stances based on similarity between user vectors and stance vectors (pro-life and pro-choice).}
\label{fig:case_study_stances}
\end{figure}

\begin{figure}[t]
\centering
\includegraphics[width=0.32\textwidth]{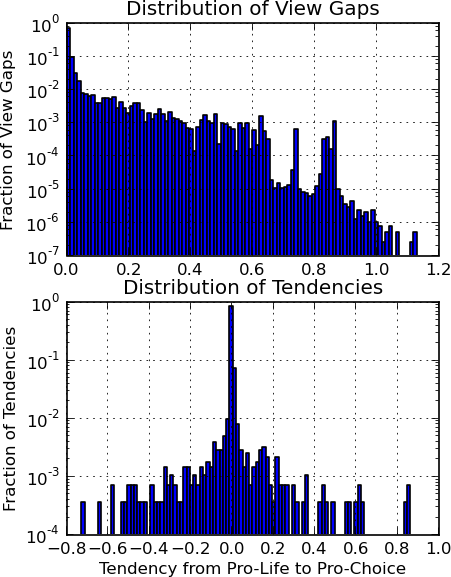}
\caption{Distribution of view gaps in abortion (distances between user views for a pair of users) and tendency to pro-life or pro-choice stances for regular people in our dataset.}
\label{fig:case_study_tendencies}
\end{figure}

We estimated the candidates' stance on abortion by computing the cosine similarity between their user vectors and the pro-life and pro-choice stance vectors. The two-dimensional stances are displayed on Figure \ref{fig:case_study_stances}. The \emph{view gaps} are defined as the distance between their stances. We also defined a \emph{candidate tendency} as the difference between their similarities with the pro-choice stance vector and the pro-life vector. Figure \ref{fig:case_study_tendencies} shows the distribution of view gaps (top) and tendencies (bottom). The average view gap is $0.034 \pm 0.096$, whereas the median is $0.0051$ and the maximum is $1.129$. The average tendency is $0.00078 \pm 0.07245$, while the median is $0.00052$. Based on these results, our candidate pool is equally distributed between pro-life and pro-choice users, and most of the view gaps differences are small.

\subsection{Topical Diversity}

\begin{figure}[htb]
\centering
\includegraphics[width=0.34\textwidth]{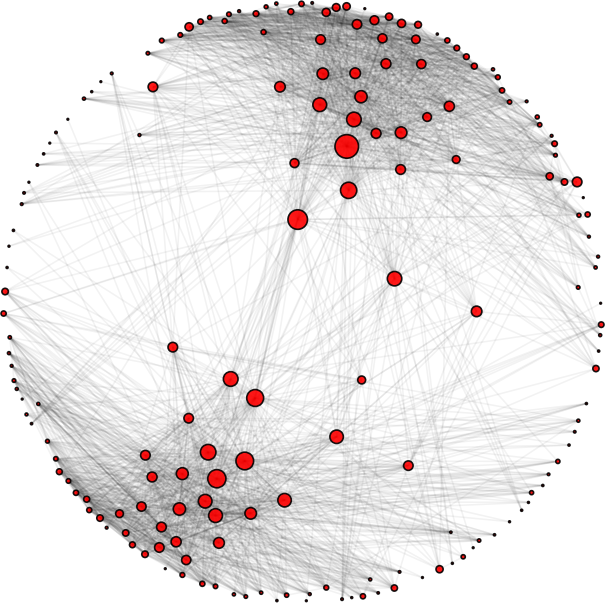}
\caption{\emph{Topic Graph}, where each node is a topic obtained with LDA applied to our corpus of user documents in Chile. Node size encodes \emph{betweenness centrality}. Two nodes are connected if both topics contribute to one user document. Only the upper-decile of edges in terms of number of user documents is kept on the graph.}
\label{fig:case_study_topic_graph}
\end{figure}

To verify that \emph{user preferences} can be estimated from our candidate pool, we explore the topical diversity of the users using topic modelling. We applied \emph{Latent Dirichlet Allocation} \cite{blei2003latent} to a corpus where each document contains candidate tweets, after filtering the top-50 corpus specific stopwords (in a similar way to \cite{ramage2010characterizing}). LDA is a generative model that, given a number of topics $k$ and a corpus of documents, estimates which words contribute to each latent topic. Then, we can estimate what is the probability for a latent topic to contribute to an arbitrary document. 

We ran LDA with $k = 300$. Then we built an undirected \emph{topic graph} where each topic is a node, and two nodes are connected if the two corresponding topics contribute to the same \emph{candidate tweets document}. The edge weight is the fraction of users that contributed to the edge. Having an almost fully connected graph, we filter the edges based on their weight, leaving only those in the upper-decile. After removing the disconnected nodes we have $181$ nodes. Finally, we compute the \emph{betweenness centrality} \cite{newman2005measure} for all remaining nodes to find intermediary topics. This graph is shown in Figure \ref{fig:case_study_topic_graph}, where two topical communities can be seen. The most descriptive terms for the top-5 intermediary topics are shown in Table \ref{table:topic_keywords}. These terms were computed using a TF-IDF inspired term-score formula \cite{blei2009topic}.

\begin{table}[htbp]
\centering
\scriptsize
\begin{tabulary}{0.45\textwidth}{L|L}
\toprule
Topic & Keywords  \\
\midrule
$57$ &
\texttt{santiago}, \texttt{partido}, \texttt{derecha}, \texttt{parte}, \texttt{trabajo}, \texttt{nacional} \\
\midrule $156$ & 
\texttt{\#atencion}, \texttt{\#popular}, \texttt{\#familiar}, \texttt{\#2014}, \texttt{\#creativo}, \texttt{\#recomendada} \\
\midrule $10$ &
\texttt{\#junior}, \texttt{\#trabajando}, \texttt{\#cueca}, \texttt{\#resumen}, \texttt{\#cambios}, \texttt{\#facha} \\
\midrule $157$ &
\texttt{\#vende}, \texttt{\#camara}, \texttt{\#drama}, \texttt{\#marco}, \texttt{\#enfermo}, \texttt{\#asqueroso} \\
\midrule $99$ &
\texttt{\#equipos}, \texttt{\#ataque}, \texttt{\#catolicos}, \texttt{\#comunicaciones}, \texttt{\#absurdo}, \texttt{\#television} \\
\bottomrule
\end{tabulary}
\caption{Representative keywords for intermediary LDA topics.}
\label{table:topic_keywords}
\end{table}

We observe that the top intermediary topics include mostly ``neutral'' keywords; they are not related to abortion-related issues. In particular, the most descriptive words for each topic are: \texttt{santiago} (location name), \texttt{\#atencion} (a hashtag of general use), \texttt{\#junior} (referring to the soccer player \emph{Junior Fern\'andez}), \texttt{\#vende} (a hashtag to sell things) and \texttt{\#equipos} (a hashtag that refers to soccer teams). Although we did not analyze the non-intermediary topics (for space constraints) the fact that we have identified some level of neutrality in the intermediary topics indicates that it is possible to link two topical communities through trivial things such as sports and item exchange.

\section{User Evaluation}

In this section we evaluate with a user study our data portrait paradigm, and using our methodology to defining user preferences, stances, etc, described in the previous section.

\spara{Participants.}
Participants were recruited from social networks. No compensation was offered. We recruited $37$ participants ($26$ male and $11$ female), all of them from Chile and active users of Twitter. When participants were asked to rate their experience with social networks they scored themselves $3.84 \pm 0.80$ in average using a Likert scale from $1$ to $5$. In addition,  $84\%$ ($31$) of participants reported to have accepted recommendations from other persons, whereas $54\%$ ($20$) reported to have accepted recommendations from Twitter \emph{who to follow}. When asked in the post-study survey for their position on abortion, $86\%$ ($32$) support the legalization of abortion.

\begin{figure}[htb]
\centering
\includegraphics[width=0.46\textwidth]{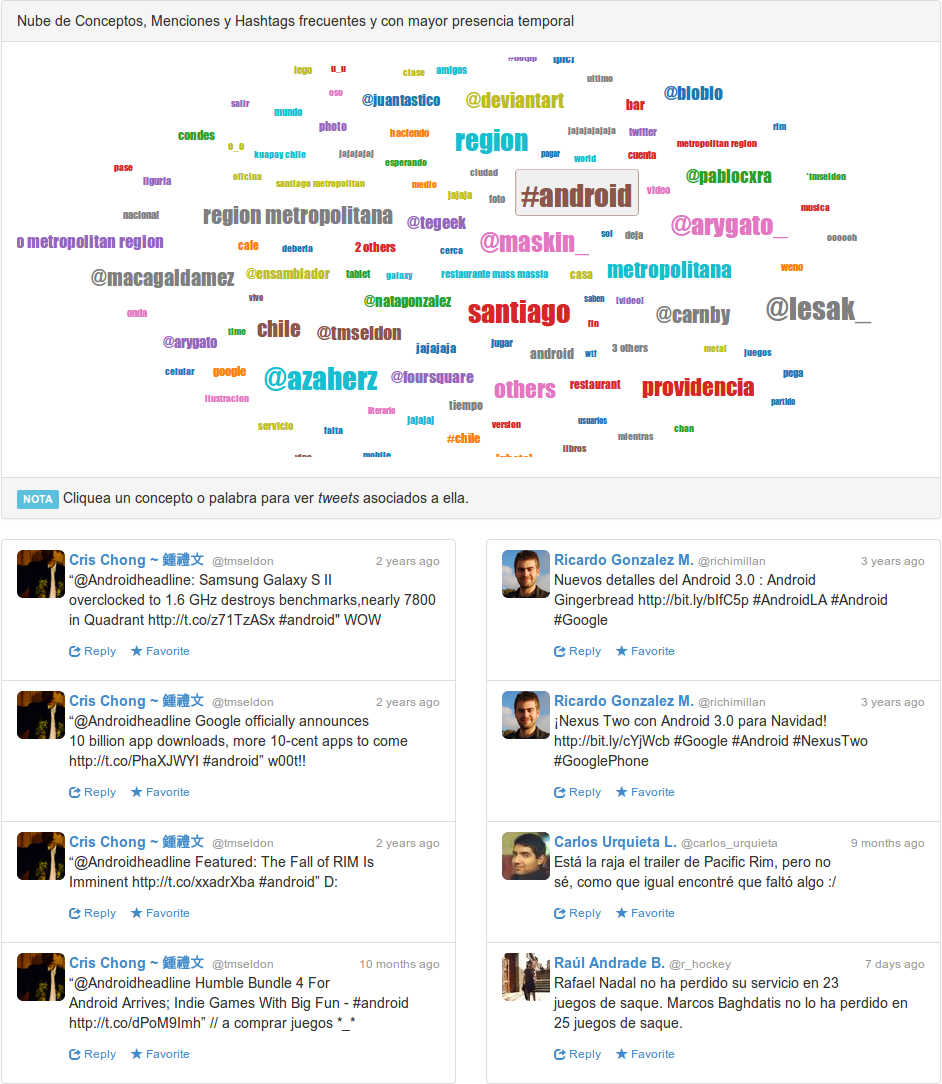}
\caption{Baseline interface using a wordcloud and standard lists of tweets.}
\label{fig:app_baseline}
\end{figure}

\spara{Apparatus.}
For each participant we crawled their last $3,200$ tweets (if applicable), which is the limit imposed by the Twitter API. Then we estimated their \emph{user preferences}  and \emph{user views} on abortion according to our methodology. 
Participants were tested in an online-setting. Our data portrait design (see Figure \ref{fig:application}) and the baseline design (see Figure \ref{fig:app_baseline}) were implemented in HTML and Javascript using \texttt{d3.js} \cite{bostock2011d3}. Before the start of the experiment, we explained that we have built a visually explorable characterization of them and that we will display related tweets to their characterization. Each participant was given a unique url with their portrait to visit. After the experiment, participants filled a post-study survey with two parts: one part contained usability questions and a second part with  questions related to the sensitive issue of our case study.
Participants were not told that the recommendations they would receive were from people with eventually opposing views (if applicable), and they were not told about the sensitive issues aspect of the experiment until the second part of the post-study survey. A comment was added at the end of the experiment to explain why we were asking about abortion and sensitive issues.

\spara{Design and Procedure.}
The experiment used a between-groups design. We defined three groups: baseline, treatment I and treatment II. The baseline interface contains a wordcloud and lists of tweets with a similar format to the current Twitter interface (see Figure \ref{fig:app_baseline}), and  the recommendations consider \emph{user preferences} only. Treatment I uses the same baseline interface, but recommendations depend on both  \emph{user preferences} and \emph{view gaps} with authors with respect to abortion, as defined in our methodology. Treatment II considers both \emph{user preferences} and \emph{view gaps}, and uses our data portrait design.

\spara{Task.}
We asked participants to visit their portrait during three consecutive days, and to browse their portraits for as long as they want, but for a minimum of three minutes. They were encouraged to explore their \emph{user preferences}, but we did not explicitly encourage them to interact with others. If participants tweeted during the days of experiment, their portraits were updated. 

\begin{figure}[t]
\centering
\includegraphics[width=0.33\textwidth]{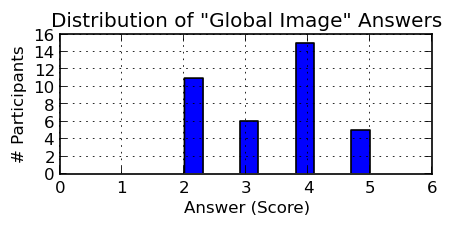}
\caption{Distribution of answers to the question \emph{Do you think the portrait presents a ``global image'' of your profile?}}
\label{fig:global_image_distribution}
\end{figure}
\begin{table*}[hbt]
\scriptsize
\begin{tabulary}{\textwidth}{L|CCC}
\toprule
\null & Control ($N = 12$) & Treatment I ($N = 12$) & Treatment II ($N = 13$) \\
\midrule
How much did you enjoy using the application? & $3.92 \pm 0.79$ & $3.50 \pm 0.90$ & $3.92 \pm 0.76$ \\
Would you use the application if it was integrated in Twitter? & $3.67 \pm 0.65$ & $3.17 \pm 1.03$ & $3.31 \pm 0.75$ \\
How much did you feel represented by the portrait? & $3.75 \pm 0.62$ & $3.08 \pm 0.51$*** & $3.69 \pm 0.95$* \\
Do you think the portrait presents a ``global image'' of your profile? & $3.42 \pm 1.16$ & $3.17 \pm 1.11$ & $3.54 \pm 0.97$ \\
Do you think the portrait allows you to discover patterns in your behavior? & $3.83 \pm 0.94$ & $3.50 \pm 1.24$ & $3.85 \pm 1.07$ \\
\midrule
Recommendation Similarity & $2.58 \pm 1.00$ & $2.58 \pm 0.90$ & $2.31 \pm 0.95$ \\
Recommendation Interestingness & $2.50 \pm 1.31$ & $3.17 \pm 1.03$ & $2.46 \pm 0.97$* \\
Recommendation Serendipity & $2.83 \pm 0.83$ & $3.08 \pm 1.08$ & $2.92 \pm 1.04$ \\
\bottomrule
\end{tabulary}
\caption{Means and standard deviations considering control group, treatment I (recommendations considering view gaps) and treatment II (organic visualization and recommendations considering view gaps). ***: $p < 0.01$. **: $p < 0.05$. *: $p < 0.1$.}
\label{table:results_groups}
\end{table*}

\subsection{Quantitative Results}
The task and measures we made were designed to evaluate if users reacted well to our data portrait paradigm. Since our design presents a non typical design, it could result into a disrupting user experience for those users who are used to the more traditional Twitter user interface. Therefore, our focus was on testing how comparable the user experiences were. 
Considering the previous statement, our hypothesis is: \emph{participants in Treatment II will have a comparable user experience than participants in Treatment I}. Table \ref{table:results_groups} shows the average and standard deviations of answers given by participants on the post-study survey. Participants answered using a Likert scale from $1$ to $5$. The dfferences in means were tested with \emph{Welch's t-test}. The table also lists the questions that were asked.\footnote{All questions were asked in Spanish. We translate them in English for ease of understanding. Note that ``\emph{serendipity in recommendations}'' was translated from \emph{``surprise in recommendations''}, as \emph{serendipity} does not exist in Spanish. Every other relevant word in our context has been directly translated.}

The results support our hypothesis. The scores obtained for the questions \emph{How much did you enjoy using the application?} and \emph{Would you use the application if it was integrated in Twitter?} are positive and do not show statistically significant differences between the groups. However, there is an unexpected finding; 
participants in treatment I felt less represented by their portraits (significant at $p < 0.01$), whereas participants in treatment II felt equally represented as those in the control group (the difference between treatment II and baseline is not significant, and the difference between treatments is almost significant at $p < 0.1$). The implications of these results are discussed in the next section.

\spara{Individual Differences.}
When looking at the distribution of answers, the answers to the question \emph{Do you think the portrait presents a ``global image'' of your profile?} did not follow a normal distribution. This is shown in Figure \ref{fig:global_image_distribution}. We separated participants in two groups, those with positive scores ($answer > 3$) and those with neutral or negative scores ($answer \le 3$). In both groups, we considered participants from the three experimental groups, as there are no significant differences in the score to the global image question between them (see Table \ref{table:results_groups}). Participants distribute equally with respect to the experimental groups across the two groups based on the global image score, as shown in Table \ref{table:score_groups}. In addition of the expected differences given the separation of groups, the high global image score group found the recommendations more serendipitous (difference is significant at $p < 0.05$) than participants in the low score group. The means and standard deviations of answers for both groups are displayed in Table \ref{table:score_group_differences}.

\begin{table}[t]
\scriptsize
\begin{tabulary}{0.45\textwidth}{L|CCC}
\toprule
Group & Cont. & Treat. I & Treat. II \\
\midrule
High Global Image Score & $7$ & $6$ & $7$ \\
Low Global Image Score & $5$ & $6$ & $6$ \\
\midrule
Has Abortion-Related Tweets & $7$ & $7$ & $6$ \\
No Abortion-Related Tweets & $5$ & $5$ & $7$ \\
\bottomrule
\end{tabulary}
\caption{Composition of participant groups for the individual differences found.}
\label{table:score_groups}
\end{table}

\begin{table*}[hbt]
\scriptsize
\begin{tabulary}{\textwidth}{L|CC}
\toprule
\null & High Global Image Score ($N = 20$) & Low Global Image Score ($N = 17$) \\
\midrule
How much did you enjoy using the application? & $3.95 \pm 0.83$ & $3.59 \pm 0.80$ \\
Would you use the application if it was integrated in Twitter? & $3.60 \pm 0.82$* & $3.12 \pm 0.78$ \\
How much did you feel represented by the portrait? & $3.90 \pm 0.64$*** & $3.06 \pm 0.66$ \\
Do you think the portrait presents a ``global image'' of your profile? & $4.25 \pm 0.44$*** & $2.35 \pm 0.49$ \\
Do you think the portrait allows you to discover patterns in your behavior? & $4.05 \pm 0.89$** & $3.35 \pm 1.17$ \\
\midrule
Recommendation Similarity & $2.35 \pm 0.93$ & $2.65 \pm 0.93$ \\
Recommendation Interestingness & $2.60 \pm 1.14$ & $2.82 \pm 1.13$ \\
Recommendation Serendipity & $3.30 \pm 0.86$** & $2.53 \pm 0.94$ \\
\bottomrule
\end{tabulary}
\caption{Means and standard deviations for two participant groups based on the score given to the question \emph{``Do you think the portrait presents a ``global image'' of your profile?''}  ***: $p < 0.01$. **: $p < 0.05$. *: $p < 0.1$.}
\label{table:score_group_differences}
\end{table*} 
\begin{table*}[hbt]
\scriptsize
\begin{tabulary}{\textwidth}{L|CC}
\toprule
\null & Has Abortion-Related Tweets ($N = 13$) & No Abortion-Related Tweets ($N = 12$) \\
\midrule
How much did you enjoy using the application? & $3.85 \pm 0.80$ & $3.58 \pm 0.90$ \\
Would you use the application if it was integrated in Twitter? & $3.38 \pm 0.87$ & $3.08 \pm 0.90$ \\
How much did you feel represented by the portrait? & $3.46 \pm 0.78$ & $3.33 \pm 0.89$ \\
Do you think the portrait presents a ``global image'' of your profile? & $3.38 \pm 0.96$ & $3.33 \pm 1.15$ \\
Do you think the portrait allows you to discover patterns in your behavior? & $3.54 \pm 1.05$ & $3.83 \pm 1.27$ \\
\midrule
Recommendation Similarity & $2.92 \pm 0.64$*** & $1.92 \pm 0.90$ \\
Recommendation Interestingness & $3.38 \pm 0.77$*** & $2.17 \pm 0.94$ \\
Recommendation Serendipity & $3.31 \pm 1.03$ & $2.67 \pm 0.98$ \\
\bottomrule
\end{tabulary}
\caption{Means and standard deviations considering participants who have tweeted (or not) about abortion in the past and have received diverse recommendations in respect to abortion. ***: $p < 0.01$.}
\label{table:score_tweeted_about_abortion}
\end{table*}

In the post-study survey we asked if participants have tweeted about abortion before. Based on their answers, we separated participants who received recommendations from people with opposing views  into two groups: \emph{Have tweeted about abortion before} ($N = 13$ of a total of $20$) and \emph{No abortion related tweets} ($N = 12$ of a total of $17$). We selected only participants from our treatments because we are interested in understanding if previous behavior relate with their user experience. The means and standard deviations of the answers are displayed in Table~\ref{table:score_tweeted_about_abortion}. In terms of enjoyment and data portrait identification there are no significant differences. However, participants who have tweeted before about abortion and received recommendations from people with opposing view on abortion gave a higher similarity and interestingness score to the recommendations than those who did not (differences are significant at $p < 0.01$)\footnote{Note that those differences hold the same significance level if we do not discriminate users who received recommendations without considering view gaps. We do not show the full results here because of space constraint.}.

\subsection{Qualitative Feedback}
We included open questions in the post-study survey to understand the responses to the questions asking to rate various aspects of our tool, as well as user views about our data portrait paradigm.\footnote{The answers to the open questions have been translated from Spanish to English.}

\spara{User Experience and Expectations.}
In the answers to the open question \emph{``How would you describe the application you have used?''}, the high rating obtained with respect to enjoyment is also reflected in the participant responses (we use P$i$ to refer to participant $i$): 
\emph{``I liked the way in which you select the points when you click on a word. I also liked a lot the colors and the tag cloud''} [\texttt{P10}, treat. II], 
\emph{``Didactic, fun and colorful''} [\texttt{P14}, treat. II],
\emph{``Friendly. Clear in terms of concepts and visual representation of the information''} [\texttt{P19}, treat. II],
\emph{``I like the connections between tweets based on keywords. It is useful for people that curates their content. I also liked the relations with other users.''} [\texttt{P24}, treat. II],
\emph{``It is a novel idea. At the beginning I did not understand how it worked, but after a couple of clicks I managed to find the `rythm'''} [P29, treat. II].

Responses to the open questions contained several suggestions. Some participants wrote that the proposed design could be improved by having a time-based filter: 
\emph{``I would add the option to filter by time. For instance, to visualize the same things 1 year ago, 2 years ago, etc.''} [\texttt{P5}, treat. II]. 
With respect to interacting with their data portrait, several participants stated they were not interested in doing so, for example, because 
\emph{``I would evaluate if mentions should appear on the application. It is good that the cloud contains topics, but I'm not convinced about mentions.''} [\texttt{P17}, treat. II]. 
There was also some doubt about the usefulness of the tool:
\emph{``Interesting, but not dynamic enough, too static to take a real benefit from it''} [\texttt{P3}, baseline],
\emph{``An interesting ``gimmick'' but not necessarily useful for the typical user I know from Twitter''} [\texttt{P11}, treat. I],
\emph{``Pretty, but not so useful''} [\texttt{P18}, treat. II]. This was expected, as \emph{casual infovis} systems \cite{pousman2007casual} are not there to solve a task, and as such can be considered as not very useful. 
However, our aim was to define a paradigm with no task in mind, apart of just ``browsing''.

\spara{Discoveries.}
As in previous work \cite{viegas2006visualizing}, and as expected, many users discovered something about themselves:
\emph{``The cloud shows many curious terms that sometimes you do not notice how frequently you use them''} [\texttt{P2}, baseline], 
\emph{``I did not know that I wish so many happy birthdays in Twitter''}  [\texttt{P12}, treat. I],
\emph{``I found some tweets I did not even remember I had written''} [\texttt{P15}, treat. I],
\emph{``I was surprised by the most highlighted concept, it was a discovery. I knew it was important but not that it was the most. Really good finding''} [\texttt{P23}, treat. I],
\emph{``I was surprised by the amount of tweets associated to certain concepts I did not consider I was using them so much, but here they were exposed. I liked it because it helps you to understand your profile''} [\texttt{P24}, treat. II].
But some users did not discover anything:
\emph{``Sometimes I see too many generic verbs, like `calling' or `doing'. I got dispersed results without any pattern''} [\texttt{P37}, treat. I], whereas others noted that their \emph{user preferences} were too old: 
\emph{``There were extremely old tweets (+4 years). This devalues the application, because in general my usage of Twitter is `now'''} [\texttt{P29}]. So time is important (as reported earlier a filtering according to time was suggested)
as well as finding ``interesting'' words, where interestingness remains to be defined.

\spara{Recommendations.}
Some participants accepted the recommendations made to them:
\emph{``I followed a couple of users with similar social/political opinions.''} [\texttt{P33}, treat. I], 
\emph{``I have followed only those who are similar to what I have tweeted about\ldots''} [\texttt{P37}, treat. I]. 
Other users felt that the recommendations were good, but did not follow them:
\emph{``I did not interact with anyone, but now that I think about it, there were at least a couple of tweets I should have \emph{favorited}.''} [\texttt{P27}, treat. I], 
\emph{``Effectively, I discovered something new. I did not follow them [the authors], but that is another thing, in general I do not follow many people because I want to keep my timeline clean. But I did consider following new people\ldots''} [\texttt{P29}, treat. II]. This is not too surprising as our evaluation was not a longitudinal study, as expecting all good recommendations to be followed in an one-off study is unrealistic.  

The recommendations however were not always good. Their quality varied with respect to the \emph{user preferences}: 
\emph{``I had the feeling that precision in recommendation was greater for central terms. They become more imprecise for the rest.''} [\texttt{P27}, treat. I].
Some users stated that the recommendations were too old to be of interest: 
\emph{``This tool is useful to find people, but not to RT or FAV older stuff''} [\texttt{P6}, treat. I]. 
Finally some recommendations were just plain wrong, mostly because the meaning of a word was wrongly
interpreted (the vector space model as implemented here does now distinguish between meanings):
\emph{``Recommendations were similar syntactically but not semantically [...]. They were interesting, in the sense of seeing how others use the same words in different contexts, but because of that the other users were not topically relevant for me. It was surprising, but not in the sense of `ah, this person is writing about the same things'.''} [\texttt{P19}, treat. II],
\emph{``The recommendations had no relation. I mean, they had words in common, but not content!''} [\texttt{P24}, treat. II]. It will be important in future work, and when carrying out a longitudinal study that we improve the quality of the recommendations.

\section{Discussion}

\spara{Theoretical Implications.}
In the study reported in~\cite{faridani2010opinion}, using visualization was shown to affect the behavior of users discussing sensitive issues, but it had no impact on selective exposure. Other works applying incremental changes to the user interfaces did not lead to any change in user behaviour, but helped understanding how users behaved and the limitations of current trends in user interfaces. 
With this respect, our visualization approach was different: while our design could be described as non-traditional (as it is based on the use of organic layout), we started with something that is familiar to users (wordclouds) and added an organic element that enabled a different way for users to interact with the wordcloud but \emph{in a manner} close to what users expect with wordclouds.
We believe that incremental modifications of a user interface can help preventing resistance to change, as it is well-known that significant changes in social networks and online-services can generate some  amount of user anger and frustration.
Indeed, it has be noted that \emph{``one argument for deliberately designing evocative visualizations for online social environments is the existing default textual interfaces are themselves evocative, they simply evoke an aura of business-like monotony rather than the lively social scene that actually exists''} \cite{donath2002semantic}. 

Our results support our claims.
First, if we consider identification with the data portrait as an emotional reaction, we found that including diverse recommendations with respect to user views degraded the emotional response, whereas using our data portrait design helped to recover it and put it back to the same level as before.
Second, there were no differences in enjoyment between the different experimental groups; overall enjoyment was high. Thus, while we expected resistance to change, we found that our data portrait design did not cause any resistance.
Hence, our results open a path in \emph{the usage of organic visualization design to revert the effect of potentially discomfortable information}. Instead of diverse recommendations with respect to user views, other types of discomfortable (e.g.~propaganda) or annoying information (e.g.~unwanted advertising) could lead to similar desirable effect with our organic data portrait design. 

\spara{Practical Implications.}
Our results indicate that \emph{individual differences matter}. Indeed, the rating of the three aspects of the recommendations we measured (interestingness, similarity and serendipity) were related to individual factors: users who tweeted about a sensitive issue gave higher scores for interestingness and similarity compared to those who did not. Also, people who felt represented by the portrait gave higher score to serendipity. This has two main implications:
\begin{enumerate}\itemsep0pt
	\item \emph{Users who openly speak about sensitive issues are more open to receive recommendations authored by people with opposing views}. Because \emph{openness} of Twitter users can be predicted \cite{quercia2011our}, these individual differences can  be detected without explicitly having to determine whether a user is concerned about specific sensitive issues.
	\item \emph{Using the data portrait is a signal that users strive for serendipity}. The serendipity rating of the recommendations is related with how users felt represented by their portrait, and those users that gave a statistically significant higher score to the question: \emph{``Would you use the application it if was integrated in Twitter?''}. Therefore, if a data portrait was to be included in Twitter, the fact that a user is interacting with it could be a strong indication of his or her willingness to value serendipitous recommendations.
\end{enumerate}

\spara{Abortion in Chile.}
In Section \ref{section:case_study}, we studied the distribution of user views on abortion in Chile. Although Chile is a highly polarized country on a number of sensitive issues, and in particular abortion, we saw that user tendencies towards pro-life and pro-choice stances are equally distributed, and polarity is rather low (see Figure \ref{fig:case_study_tendencies}). 
How these observations correspond to the reality in Chile is hard to determine, as several public opinion surveys report contradictory results with respect to this issue \cite{abortionChile}. Nonetheless, we expected a higher polarization.
A possible explanation is that our TF-IDF weighting used in the \emph{stance vectors} is softening the stronger stances on the issue.
It will be important to revisit the computation of the stance vectors, for example, not only by considering the words that are deemed to relate to the sensitive issue under study but how these words are used by the users.   
A second explanation is that users may be less vocal about this (or other) sensitive issue than our expectations. Previous work demonstrated that the ``average'' users are not as vocal as politically involved people \cite{ICWSM136128}. 
This somewhat reflects what has been reported in the Chilean press about users and abortion: pro-life and pro-choice people rarely interact with each other (so no need to become vocal) and most of what they do is to \emph{retweet} agreeable information \cite{abortionTwitterChile}. Considering these retweets in our TF-IDF weighting used to generate the stance vectors may have shown stronger polarization. We did not consider retweets merely to obtain a user stance based on his or her own tweets. There is thus room for improvement here. Nonetheless, to the extent of our knowledge this is the first work looking into linking people with opposite view on a given sensitive issue, and our main focus was the usage of the data portrait paradigm, and to show its potential towards our goal. 

\spara{Limitations}
We discussed that the quality of the recommendations will require improvement. Although this may have an impact on our results, we believe that overall the recommendations were acceptable to inform us about the potential of the proposed data portrait paradigm as a way to link people having opposite views on sensitive issues.

\spara{Future Work}
In addition to test better recommendation algorithms, individual differences and ways to detect and measure them are two important problems to solve in future work. Finally, to evaluate our methodology and our data portrait design, a longitudinal study is required to establish whether people with opposing views become connected, for how long and why (or why not), and the overall effect on their beliefs. 

\section{Conclusions}

We presented a methodology to determine user preferences and user views about sensitive issues, and to use those to recommend relevant content authored by others who have opposite views. This information was presented through a data portrait with an organic visual design. Hence, our approach is different from previous work in that we propose an indirect way to connect dissimilar people. 
We evaluated our design and methodology through a case study. We analyzed how users in Chile tweeted about abortion, and used the outcomes to perform a usability study with Chilean users using a proof-of-concept implementation of our recommendation approach in a data portrait paradigm.

We found that using an organic visualization design has an impact on how users felt identified by the portrait. 
In addition, we found relationships between individual differences and recommendation ratings: openness in tweeting about sensitive issues is related to higher interestingness and similarity to user preferences, whereas serendipitous recommendation is related to higher identification with the data portrait. These results have implications for the design of recommender systems and the usage of visualization in social networks. 
We conclude that an indirect approach to connecting people with opposing views has great potential and using an organic design constitutes a first step in that direction, where so-called \emph{view gaps} could be more tolerated.

{
\clearpage
\bibliographystyle{plain}
\bibliography{virtual_portraits}
}

\end{document}